\documentclass[baaa]{baaa}

\usepackage[pdftex]{hyperref}
\usepackage{subfigure}
\usepackage{natbib}
\usepackage{helvet}
\usepackage[font=small]{caption}

\begin{document}


\journalvol{57}
\journalyear{2014}
\journaleditors{A.C. Rovero, C. Beaugé, L.J. Pellizza \& M. Lares}


\contriblanguage{0}


\contribtype{3}

\thematicarea{3}

\title{Astrosismolog\'ia de estrellas enanas blancas}


\titlerunning{Astrosismolog\'ia de enanas blancas}


\author{A.H. C\'orsico\inst{1,2}}
\authorrunning{C\'orsico}
\contact{acorsico@fcaglp.unlp.edu.ar}

\institute{Facultad de Ciencias Astron\'omicas y Geof\'isicas, 
Universidad Nacional de La Plata  \and
  Instituto de Astrof\'isica La Plata (IALP - CONICET)
}


\resumen{La mayor\'ia de las estrellas de masa baja e intermedia 
que pueblan nuestro Universo finalizar\'an sus vidas como estrellas 
enanas blancas. Estos antiguos remanentes estelares guardan en su interior un 
valioso registro de la historia evolutiva de sus estrellas progenitoras,
proveyendo abundante informaci\'on acerca de la formaci\'on y 
evoluci\'on de las estrellas, asi como tambi\'en una estimaci\'on de 
la edad de una variedad de poblaciones 
estelares, tales como nuestra Galaxia,  c\'umulos abiertos y globulares.
Aunque la composici\'on qu\'imica, 
la temperatura y la gravedad superficiales de las enanas blancas pueden ser 
inferidas a partir de la espectroscop\'ia, la estructura interna de estas 
estrellas compactas puede ser revelada solo por medio de la 
astrosismolog\'ia, una t\'ecnica basada en la comparaci\'on entre 
los per\'iodos observados en estrellas variables pulsantes y 
los per\'iodos te\'oricos calculados sobre modelos estelares 
apropiados. En este informe,  primero describimos brevemente las
propiedades b\'asicas de las estrellas enanas blancas y las distintas 
familias de enanas blancas pulsantes conocidas, 
y a continuaci\'on presentamos dos de los \'ultimos estudios
realizados por el Grupo de Evoluci\'on Estelar y Pulsaciones de La Plata
en este activo campo de investigaci\'on.}

\abstract{Most of low- and intermediate-mass stars that populate the Universe
will end their lives as white dwarf stars.  These ancient stellar
remnants have encrypted inside a precious record of the evolutionary
history of the progenitor stars, providing a wealth of information
about the evolution of stars, star formation, and the age of a variety
of stellar populations, such as our Galaxy and open and globular
clusters. While some information  like surface chemical composition,
temperature and gravity of white dwarfs can be inferred from
spectroscopy, the internal structure of these compact stars can be
unveiled only by means of asteroseismology, an approach based on the
comparison between the observed pulsation periods of variable stars
and the periods of appropriate theoretical models. In this  communication, we first
briefly describe the physical properties of white dwarf stars and the various  
families of pulsating white dwarfs known up to the
present day, and then we present two recent  analysis carried out
by the La Plata Stellar Evolution and Pulsation Group in this
active field of research.}




\maketitle

\section{Introducción}
\label{INTRO}

Las estrellas enanas blancas constituyen el destino final de la mayor\'ia ($\sim 97 \% $) 
de todas las estrellas que pueblan el Universo, incluido nuestro Sol. 
En efecto,  todas las estrellas que en la 
Secuencia Principal (MS) poseen una masa estelar menor a $\sim 8 M_{\odot}$,
culminar\'an sus vidas pasivamente como enanas blancas. El  lector interesado en 
detalles acerca de la formaci\'on y evoluci\'on de enanas blancas puede consultar 
el art\'iculo de revisi\'on de \cite{review}. Aqu\'i solo daremos una 
breve revisi\'on de las principales caracter\'isticas de estas estrellas. 
Las enanas blancas son 
objetos compactos, caracterizados por densidades medias del orden de 
$\overline{\rho} \sim 10^6$ gr/cm$^3$ 
(en comparaci\'on, $\overline{\rho_{\sun}}= 1.41$ gr/cm$^3$) y radios 
de aproximadamente $R_{\star} \sim 0.01 R_{\sun}$, y se encuentran cubriendo un amplio
rango de temperaturas ($4000 \lesssim T_{\rm eff} \lesssim 200\,000$ K) y por lo tanto 
un amplio rango de luminosidades ($0.0001 \lesssim L_{\star}/L_{\sun}\lesssim 1000$). 
La distribuci\'on de masas de las enanas blancas alcanza un 
m\'aximo bien pronunciado en $M_{\star} \sim 0.6 M_{\sun}$, aunque el rango de masas 
es bastante amplio ($0.15 \lesssim M_{\star}/M_{\sun} \lesssim 1.2$). Las
enanas blancas con masa promedio probablemente poseen n\'ucleos de $^{12}$C y $^{16}$O, 
pero se cree que las menos masivas deber\'ian albergar n\'ucleos compuestos por $^4$He, 
y las mas masivas n\'ucleos hechos de $^{16}$O, $^{20}$Ne y $^{24}$Mg.
 
\begin{figure}[!ht]
  \centering
  \includegraphics[width=0.45\textwidth]{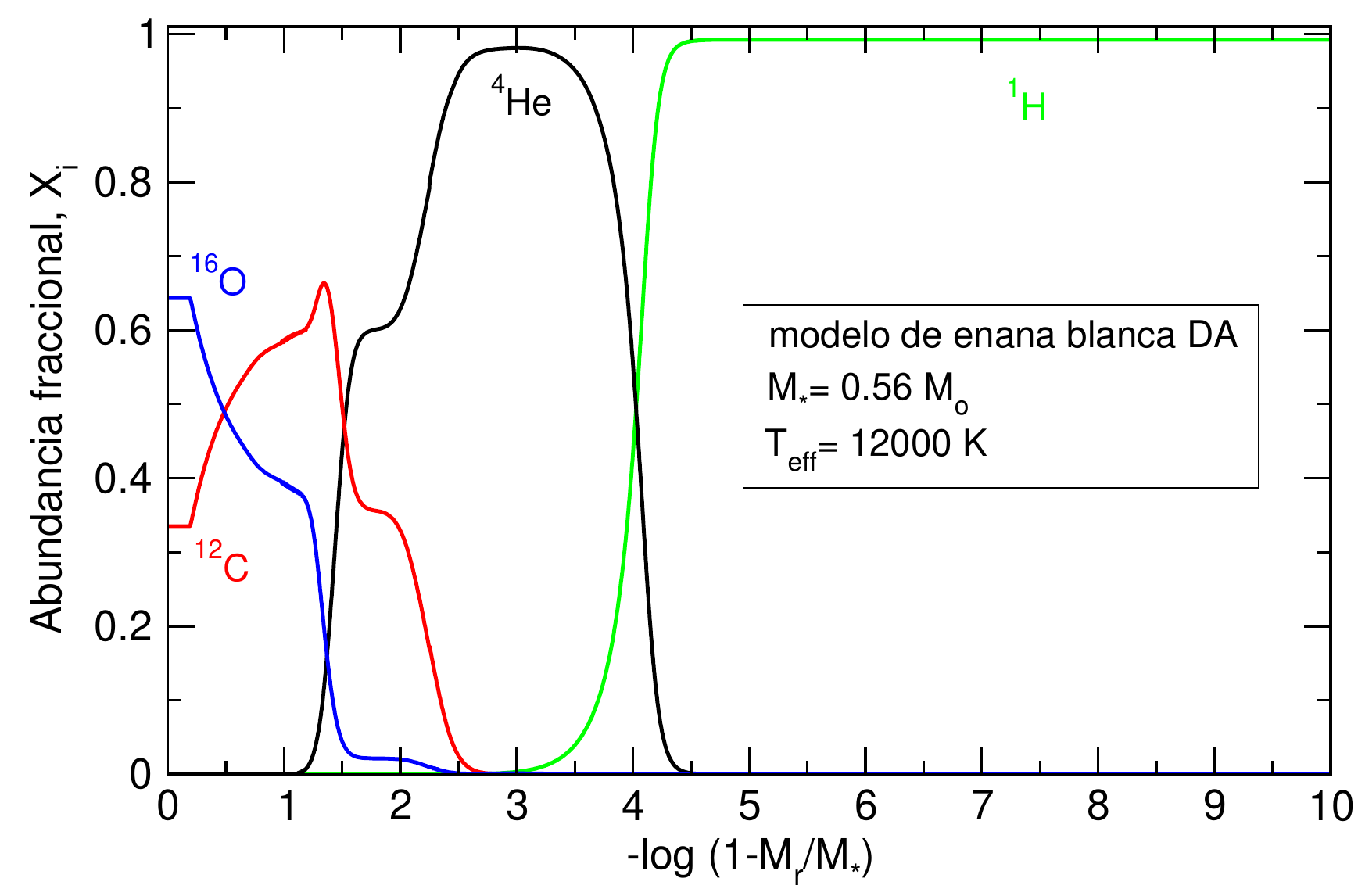}
  \caption{Estructura qu\'imica interna (abundancias fraccionales $X_i$) 
de un modelo de estrella enana blanca tipo espectral DA. El centro de 
la estrella se  localiza en $-\log(1-M_r/M_{\star})= 0$.}
  \label{fig_01}
\end{figure}

Debido a las altas densidades reinantes, la ecuaci\'on de estado que gobierna la mayor parte 
de una enana blanca es la correspondiente a  un gas de Fermi de electrones degenerados, 
los cuales aportan la mayor parte de la presi\'on. A su vez, los iones no degenerados 
contribuyen a la masa de la estrella y al contenido cal\'orico acumulado durante 
las etapas evolutivas 
previas. En t\'erminos generales, la evoluci\'on de una enana blanca consiste en un 
gradual enfriamiento, durante el cual las fuentes de energ\'ia por reacciones nucleares 
son irrelevantes. En virtud  de las enormes gravedades superficiales 
($\log g \sim 8$), 
las diferentes especies qu\'imicas
en estas estrellas se encuentran bien separadas debido al efecto de 
asentamiento gravitacional ({\em gravitational settling}).  
Seg\'un la especie qu\'imica
 dominante en la superficie de las enanas blancas, estas se clasifican en dos
grupos principales, las de tipo espectral DA ($\sim 80 \%$ del total, 
atm\'osferas ricas en H) y las de tipo espectral DB ($\sim 15 \%$ del total, 
atm\'osferas ricas en He). En la Figura \ref{fig_01} se muestra
la estructura qu\'imica estratificada de un modelo de enana blanca tipo DA. 
Como puede apreciarse, el $99 \%$ de la masa total de la estrella 
[$-\log(1-M_r/M_{\star}) \lesssim 2$] est\'a compuesto de C y O, siendo la 
masa de la capa de He de $\lesssim 0.01 M_{\star}$ y la masa de la envoltura 
de H de $\lesssim 0.0001 M_{\star}$. 

¿Por qu\'e es importante estudiar las enanas blancas? Una de las
principales razones es que son objetos muy antiguos, con edades 
del orden de $\tau \sim 10^9 - 10^{10}$ a\~nos, y por lo tanto
contienen valiosa informaci\'on acerca de la evoluci\'on de las estrellas 
desde el nacimiento hasta su muerte, y acerca de la tasa de formaci\'on estelar 
a lo largo de la historia de nuestra Galaxia. El estudio 
de las enanas blancas tiene vastas aplicaciones a varios campos de la astrof\'isica 
moderna, tales como:

\begin{itemize}

\item Estimaci\'on de la edad de poblaciones estelares tales como c\'umulos
abiertos y globulares, el disco y el halo gal\'actico, a trav\'es de la funci\'on 
luminosidad de enanas blancas \citep[``cosmocronolog\'ia''; ver, por ejemplo,][]{nature};

\item Estudios de enanas blancas como progenitoras de SNIa y variables 
catacl\'ismicas (novas): eventos energ\'eticos ($E \sim 10^{44} - 10^{51}$ erg) de 
transferencia de masa sobre una enana blanca desde su compa\~nera;

\item Investigaci\'on de fen\'omenos f\'isicos tales como la cristalizaci\'on,
las propiedades de part\'iculas fundamentales (neutrinos, axiones), 
variaci\'on de las constantes fundamentales, etc. Esto es, el uso de las 
enanas blancas como ``laboratorios c\'osmicos'', en virtud de 
las densidades y presiones extremadamente altas reinantes en los interiores de estas 
estrellas, que no pueden replicarse en laboratorios terrestres.

\end{itemize}

Tradicionalmente, las t\'ecnicas empleadas para estudiar estrellas enanas blancas 
han sido la espectroscop\'ia y fotometr\'ia, que permiten inferir la temperatura 
efectiva, gravedad superficial y composicion qu\'imica, la magnitud de los 
campos magn\'eticos (si est\'an presentes), etc. Otra t\'ecnica empleada es la 
astrometr\'ia, la cual permite derivar paralajes y distancias, e incluso la gravedad 
superficial en casos de enanas blancas fr\'ias con espectros carentes de l\'ineas. 
Una t\'ecnica mas moderna, y en muchos aspectos mas potente, es la
denominada ``astrosismolog\'ia'', que permite extraer informacion del \emph{interior}
de las enanas blancas. Los fundamentos de esta t\'ecnica son descriptos en la siguiente 
Secci\'on.

\section{Pulsaciones estelares y astrosismolog\'ia} 
\label{PULSA_ASTRO}

La astrosismolog\'ia es una disciplina basada en el siguiente principio: 
\emph{``estudiar como vibra un sistema en sus modos normales para inferir sus propiedades 
mecánicas''}. En el caso de las estrellas, las vibraciones son pulsaciones u oscilaciones 
globales que permiten ``ver'' el interior estelar, el cual es inaccesible 
a trav\'es de t\'ecnicas tradicionales. El principio b\'asico de la astrosismolog\'ia 
es la comparaci\'on de los per\'iodos de pulsaci\'on de modelos te\'oricos con los 
per\'iodos de oscilaci\'on medidos en estrellas variables pulsantes. En el caso de
las enanas blancas, la astrosismolog\'ia permite extraer informaci\'on acerca de la  
masa estelar, la estratificaci\'on qu\'imica, la composici\'on qu\'imica del n\'ucleo, la
presencia e intensidad de campos magn\'eticos, las propiedades de la rotaci\'on estelar, 
la f\'isica de la convecci\'on, etc.  Reportes muy detallados de la astrosismolog\'ia 
aplicada a estrellas enanas blancas pueden encontrarse en los art\'iculos de revisi\'on 
de \cite{2008ARA&A..46..157W} y \cite{review}. 

Las pulsaciones estelares son modos propios de las estrellas. Estos modos 
pueden pensarse como  ondas estacionarias en 3 dimensiones. Cada estrella posee 
un espectro \'unico de autofrecuencias discretas que est\'a fijado por su estructura 
interna (``frecuencias naturales''), asociadas con autofunciones que proporcionan 
las variaciones espaciales de los diferentes par\'ametros f\'isicos de la estrella 
al oscilar. En enanas blancas el espectro de frecuencias de pulsaci\'on es 
extremadamente sensible a la estratificación qu\'imica interna. Las pulsaciones estelares
pueden ser radiales, que conservan la forma esf\'erica (Cefeidas, RR Lira, Miras, etc). 
Estas son un caso particular de una clase muy general de movimientos
oscilatorios, denominados pulsaciones no radiales. Estas \'ultimas no mantienen la 
forma esférica. Pulsaciones no radiales se detectan rutinariamente en el Sol, en estrellas 
variables tipo-solar, gigantes rojas, $\delta$ Scuti, $\gamma$ Doradus, $\beta$ Cefei, 
SPB, WR, subenanas B, enanas blancas y pre-enanas blancas variables.

\begin{figure}[!ht]
  \centering
  \includegraphics[width=0.45\textwidth]{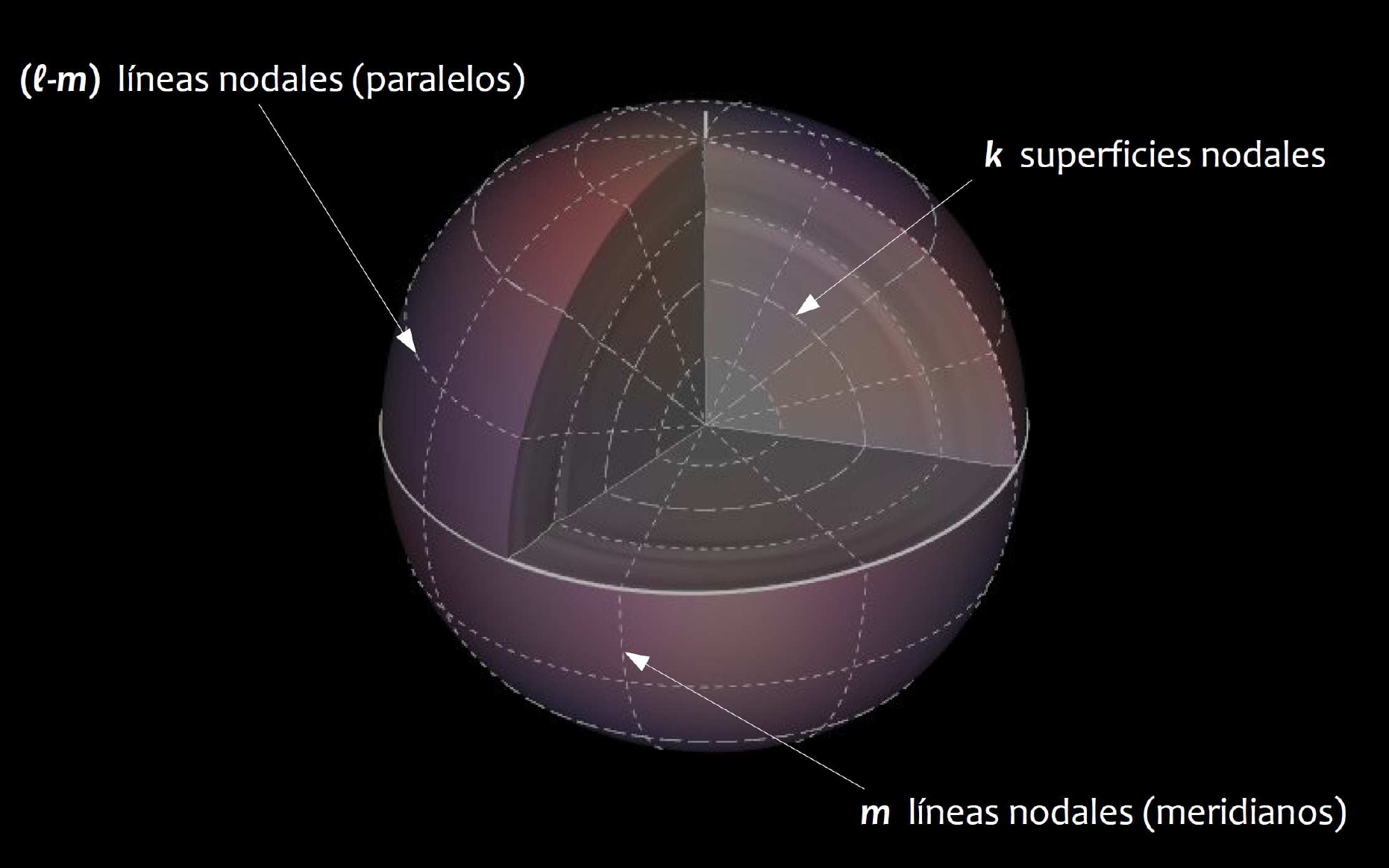}
  \caption{Esquema ilustrativo indicando el significado de los 
   n\'umeros cu\'anticos $(\ell, m, k)$ que caracterizan a un modo de 
   pulsaci\'on no radial (adaptado 
de  http://userpages.irap.omp.eu/$\sim$scharpinet/glpulse3d).}
  \label{fig_02}
\end{figure}

En el marco de la teor\'ia lineal (peque\~nas amplitudes de pulsaci\'on), las 
deformaciones de una estrella al oscilar en modos esferoidales est\'an especificadas por 
el vector desplazamiento Lagrangiano \citep{1989nos..book.....U}:

\begin{equation} 
\vec{\xi}_{n \ell m}= \left(\vec{\xi}_r, \vec{\xi}_{\theta}, 
\vec{\xi}_{\phi} \right)_{n \ell m}
\end{equation}

\noindent cuyas componentes en coordenadas esf\'ericas son:

\begin{eqnarray}
\vec{\xi}_r & = & \xi_r(r) Y^m_{\ell}(\theta, \phi) e^{i \sigma t} \vec{e}_r \\
\vec{\xi}_{\theta} & = & \xi_h(r) \frac{\partial Y^m_{\ell}}{\partial \theta} e^{i \sigma t} \vec{e}_{\theta} \\
\vec{\xi}_{\phi} & = & \xi_h(r) \frac{1}{\sin \theta} \frac{\partial Y^m_{\ell}}{\partial \phi} e^{i \sigma t} \vec{e}_{\phi}  \\
\nonumber
\end{eqnarray}

\noindent donde $ Y^m_{\ell}(\theta, \phi)$ son los arm\'onicos esf\'ericos, $\sigma$
es la frecuencia de pulsaci\'on, y  $\xi_r(r)$ y  $\xi_h(r)$ son las autofunciones
radial y horizontal, respectivamente. De esta manera, los modos propios tienen 
una dependencia temporal sinusoidal, una dependencia angular por medio de los 
arm\'onicos esf\'ericos, y una dependencia radial dada a trav\'es de las autofunciones,
las cuales ineludiblemente deben ser obtenidas (para modelos estelares realistas) 
a trav\'es de la resoluci\'on num\'erica de las ecuaciones diferenciales de 
pulsaci\'on \citep[ver][]{1989nos..book.....U}. Los modos de pulsaci\'on est\'an 
caracterizados por tres ``n\'umeros cuánticos'': (1) grado armónico 
$\ell = 0, 1, 2, 3, \cdots, \infty$, que representa $(\ell-m)$ l\'ineas 
nodales (paralelos) sobre la superficie estelar; (2)
orden acimutal $m= -\ell, \cdots, -2, -1, 0 , +1, +2, \cdots, +\ell$,
que representa l\'ineas nodales (meridianos) sobre la superficie estelar;
y (3) orden radial $k= 0, 1, 2, 3, \cdots, \infty$, que representa
superficies esf\'ericas nodales sobre las cuales hay desplazamiento nulo.
En Figura \ref{fig_02} se muestran en forma ilustrativa las diferentes l\'ineas y 
superficies nodales para un dado modo de pulsaci\'on caracterizado por 
un set de n\'umeros cu\'anticos $(\ell, m, k)$. 

Finalizamos nuestra descripci\'on de los modos no radiales esferoidales mencionando que 
existen dos familias de modos que se distinguen de acuerdo a las fuerzas de restituci\'on
dominantes. Los modos $p$, por un lado, involucran grandes variaciones de  
presi\'on y movimientos mayormente verticales, siendo la fuerza de restituci\'on 
dominante la compresibilidad. Poseen altas frecuencias (per\'iodos cortos).
Los modos $g$, por otra parte, involucran peque\~nas variaciones de presi\'on y
desplazamientos mayormente tangenciales. En este caso la fuerza de restituci\'on 
dominante es la flotación (buoyancy). Los modos est\'an caracterizados por bajas 
frecuencias (per\'iodos largos). En los casos en que $\ell > 1$, existe una 
tercer clase de modos, los modos $f$, con caracter\'isticas intermedias entre 
modos $p$ y $g$. En general no poseen nodos en la dirección radial ($k= 0$). 

\section{Clases de enanas blancas pulsantes}
\label{CLASES}

\begin{figure}[!ht]
  \centering
  \includegraphics[width=0.45\textwidth]{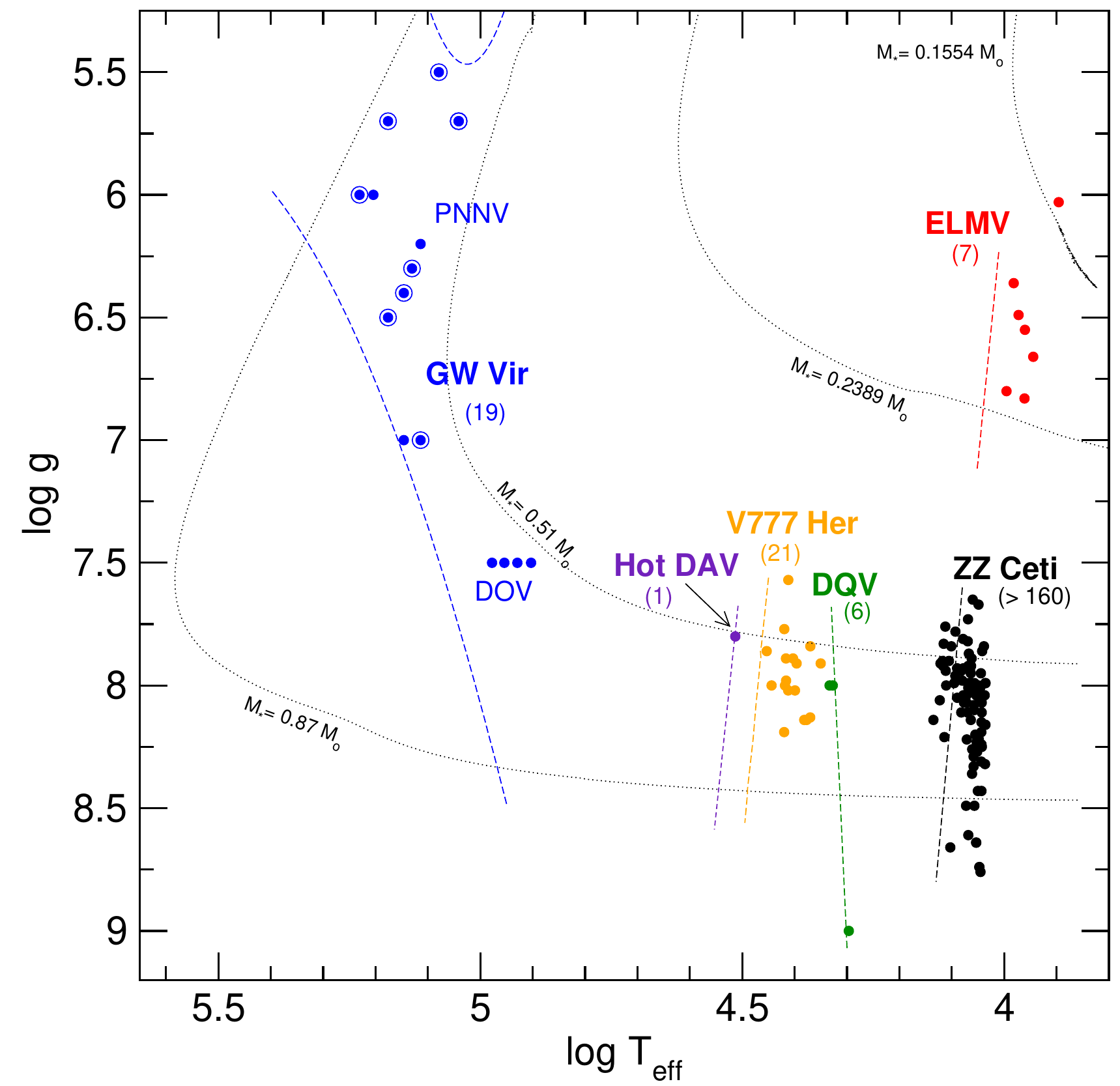}
  \caption{Localizaci\'on de las distintas clases de estrellas 
enanas blancas pulsantes en el plano $\log T_{\rm eff} - \log g$, 
indicadas con puntos de diferentes colores. Entre par\'entesis inclu\'imos 
el n\'umero de miembros conocidos hasta el momento de cada clase. Cuatro 
caminos evolutivos de modelos de enanas blancas son inclu\'idos como 
referencia con l\'ineas de puntos. Tambi\'en se muestran los borde azules 
te\'oricos de las diferentes bandas de inestabilidad con l\'ineas de trazos. C\'irculos
rodeando los puntos azules indican la presencia de una nebulosa planetaria 
(variables GW Vir PPNV).}
  \label{fig_03}
\end{figure}

\begin{table*}[!ht]
\centering
\caption{Par\'ametros estelares y propiedades pulsacionales de las distintas familias 
de enanas blancas pulsantes.}
\begin{tabular}{lccccccc}
\hline
\noalign{\smallskip}
Clase & $\#$ & $m_{\rm v}$ & $T_{\rm eff}$ & $\log g$ &  Comp. qu\'imica & Amplitudes & Per\'iodos  \\
(a\~no de descubrimiento)      &     &  (mag)     & ($\times 1000$ K)     &  (cgs)   &   superficial &  (mag)   & (s)  \\ 
\noalign{\smallskip}
\hline
\noalign{\smallskip}
GW Vir (PNNV) & 10 & $11.8-16.6$ & $110-170$  & $5.5-7.0$ & He, C, O & $0.01-0.15$ & $420-6000$ \\
(1984) &&&&&&& \\
&&&&&&& \\
GW Vir (DOV)  &  9 & $14.8-16.7$ & $70-160$   & $6.0-7.5$ & He, C, O & $0.02-0.1$  & $300-2580$ \\
(1979) &&&&&&& \\
&&&&&&& \\
Hot DAV     &  1 &    $14.4$         & $32$       & $7.5$      & H &     $0.001$        & $624$   \\
(2013) &&&&&&& \\
&&&&&&& \\
V777 Her (DBV) & 21 & $13.6-16.7$ & $22.4-29$ & $7.5-8.3$ & He (H) & $0.05-0.3$  & $120-1080$  \\
(1982) &&&&&&& \\
&&&&&&& \\
DQV            & 6  & $17.7-19.6^{(1)}$ & $19-22$ & $8-9$  & He, C & $0.005-0.015$ & $240-1100$    \\ 
(2008) &&&&&&& \\
&&&&&&& \\
ZZ Ceti (DAV) & $160$ & $12.2-16.6$ & $10.4-12.4$ & $7.5-8.75$ &  H & $0.01-0.3$ & $100-1400$  \\
&&&&&&& \\
ELMV &  7 & $16.2-18.8^{(1)}$ & $7.8-10$ & $6-6.8$ &  H & $0.002-0.044$ & $1200-6200$  \\
(2012) &&&&&&& \\
\hline
\end{tabular}
{$^{(1)}$ magnitud $g$ del sistema $ugryz$ del relevamiento SDSS.}
\label{tabla_1}
\end{table*}

A lo largo de su evoluci\'on, las enanas blancas atraviesan distintas
etapas de  inestabilidad en las cuales se tornan estrellas variables
pulsantes, exhibiendo curvas de luz con
variaciones en el \'optico y en el UV lejano. Los cambios de brillo
son debidos a modos $g$ con grado arm\'onico $\ell= 1, 2$ que producen
variaciones  en la temperatura superficial\footnote{Las variaciones en
  el radio estelar  son muy peque\~nas, del orden de $\sim 10^{-5}
  R_{\star}$, y probablemente no  contribuyen a las variaciones en el
  brillo.}. La primer enana blanca  variable fue
descubierta por \cite{1968ApJ...153..151L}. Desde entonces se han descubierto mas de
200 objetos pulsantes a trav\'es de observaciones desde Tierra, en su
gran mayor\'ia  extra\'idas del SDSS ({\em Sloan Digital Sky Survey}),  y en
los \'ultimos a\~nos tambi\'en  por medio de misiones espaciales como
la {\em Kepler Mission}\footnote{Actualmente se est\'a desarrollando
  la primera campa\~na observacional conjunta de enanas blancas
  pulsantes entre el Grupo de Evoluci\'on Estelar y Pulsaciones de La
  Plata en Argentina y  el grupo liderado por el Prof. S. O. Kepler de
  Brasil, utilizando el Telescopio de  2.15 m ``Jorge Sahade'' de CASLEO; ver
  Corti et al. (2015) en este Bolet\'in.}. Las variaciones de
brillo de estas estrellas tienen amplitudes entre 0.001 y 0.3
magnitudes. Exhiben una  gran variedad de curvas de luz, algunas
sinusoidales y de peque\~na amplitud,  otras no lineales y de gran
amplitud. Son estrellas pulsantes multimodales  (pulsan en más de un
per\'iodo), y frecuentemente exhiben arm\'onicos y  combinaciones
lineales de frecuencias que no est\'an  relacionadas con  modos
genuinos de pulsaci\'on. 

Se conocen actualmente 6 clases o familias de enanas blancas pulsantes 
(incluyendo las pre-enanas blancas pulsantes). 
En la Figura \ref{fig_03} se muestra la localizaci\'on de las 
diferentes familias en un diagrama 
$\log T_{\rm eff} - \log g$. Como referencia, hemos inclu\'ido cuatro caminos 
evolutivos de modelos de enanas blancas con masas $M_{\star}/M_{\sun}= 
0.16, 0.24, 0.51, 0.87$. Las l\'ineas de trazos indican los bordes 
calientes o ``azules'' de las bandas de inestabilidad, derivados
te\'oricamente a trav\'es de an\'alisis de estabilidad pulsacional:  
estrellas GW Vir stars \citep{2006A&A...458..259C}, estrellas hot DAV 
\citep{2013EAS....63..185S}, estrellas V777 Her \citep{2009JPhCS.172a2075C},
estrellas DQV \citep{2009A&A...506..835C}, estrellas ZZ Ceti \citep{2011ApJ...743..138G}, 
y estrellas ELMVs \citep{2013MNRAS.436.3573H}. La \'unica diferencia entre las PNNVs y las 
DOVs es que las primeras poseen una nebulosa planetaria, mientras que las \'ultimas 
carecen de la misma. 

En la Tabla \ref{tabla_1} se muestran 
en forma compacta las principales caracter\'isticas de estas familias de enanas blancas 
pulsantes. La segunda columna indica el numero de objetos conocidos a la fecha de 
escribir este reporte (noviembre de 2014), la tercer columna muestra el rango de magnitudes 
visuales aparentes, la cuarta columna incluye el rango de temperaturas efectivas 
en las cuales se las detecta (banda de inestabilidad), la quinta columna proporciona
el rango de gravedad superficial, la sexta columna indica la composici\'on qu\'imica 
superficial, la septima columna muestra las amplitues de las variaciones en las curvas 
de luz, y finalmente la octava columna proporciona el rango de per\'iodos detectados. 
Los per\'iodos de pulsaci\'on est\'an comprendidos entre
$\sim 100$ s y  $\sim 1400$ s, aunque las PNNVs y las ELMVs  exhiben
per\'iodos mucho mas largos, hasta $\sim 6200$ s. Curiosamente, los
per\'iodos de los modos $g$ de las enanas blancas son del mismo orden
de magnitud que los per\'iodos de los modos $p$ en estrellas pulsantes
no degeneradas. 

¿Cu\'al es el origen de las pulsaciones en enanas blancas? De acuerdo a los estudios
te\'oricos corrientes, los modos $g$ observados son \emph{autoexcitados} a trav\'es 
de procesos t\'ermicos\footnote{Esto, a diferencia de las pulsaciones \emph{forzadas} 
tales como la excitaci\'on estoc\'astica por convecci\'on turbulenta, en la que  modos
que son intr\'insecamente estables, son excitados por movimientos
convectivos. Este es el caso del Sol, las variables tipo solar y las 
gigantes rojas variables. 
Otro ejemplo de pulsaciones forzadas es la excitaci\'on por fuerzas de marea en sistemas 
binarios.}.  
En particular, el mecanismo $\kappa-\gamma$ involucra 
el aumento de la opacidad debido a la ionizaci\'on parcial de la especie qu\'imica 
dominante en la superficie de la estrella, que en el caso de las GW Vir son 
el C y el O, en las V777 Her es el He, en las DQV es el C, y en las ZZ Ceti y 
ELMVs es el H. Excepto en el caso de las GW Vir, las cuales carecen de convecci\'on 
superficial debido a sus altas temperaturas efectivas, 
en las otras categor\'ias un mecanismo
de excitaci\'on relacionado con la convecci\'on ({\em ``convective driving''}) juega un rol 
crucial una vez que la zona convectiva externa de la estrella
se ha profundizado lo suficiente. Finalmente, cabe mencionar el mecanismo $\varepsilon$, el 
cual es debido al efecto desestabilizante de la combusti\'on nuclear, y que
podr\'ia ser responsable  de la excitaci\'on de algunos per\'iodos cortos 
detectados en una estrella de la familia de las ELMVs (ver Secci\'on \ref{ELMVS}).

A pesar de que se conoce con alguna certeza el origen de las pulsaciones en las enanas 
blancas, poco o nada 
se sabe acerca del agente que d\'a lugar al cese de las mismas 
(borde rojo de la banda de inestabilidad). Tampoco se sabe por qu\'e este tipo de estrellas 
pulsantes (y en particular las ZZ Ceti) exhiben tan pocos per\'iodos. 
Afortunadamente, este conocimiento incompleto de la 
f\'isica de las pulsaciones en enanas blancas  no nos impide avanzar en estudios 
astrosismol\'ogicos basados en c\'alculos adiab\'aticos, en los cuales no importa el 
agente f\'isico que da origen a las pulsaciones, sino el valor de los per\'iodos en 
s\'i mismos, lo cuales dependen sensiblemente de la estructura 
interna\footnote{El sonido de las campanas no depende de \emph{c\'omo} ellas son 
tocadas \citep{1992RvMA....5..125B}.}. A continuaci\'on, nos enfocaremos en 
dos estudios te\'oricos 
recientes en los que ha estado involucrado el Grupo de Evoluci\'on Estelar y
Pulsaciones de La Plata\footnote{\tt http://fcaglp.fcaglp.unlp.edu.ar/evolgroup/} 
en el \'area de las estrellas enanas blancas pulsantes y astrosismolog\'ia.

\section{Astrosismolog\'ia de estrellas ZZ Ceti}
\label{ZZCETIS}

Las estrellas ZZ Ceti constituyen la clase mas numerosa de enanas blancas
pulsantes conocida, con mas de 160 miembros detectados hasta el momento. El gran n\'umero de 
objetos conocidos y estudiados de esta clase de estrellas variables ha 
permitido estudiar la banda de inestabilidad en forma global \citep{2006ApJ...640..956M}, 
y se ha encontrado que
\emph{todas} las enanas blancas tipo espectral DA se tornan pulsantes ZZ Ceti al atravesar
dicha banda de inestabilidad ($10\,400\lesssim T_{\rm eff} \lesssim 12\,400$ K), es 
decir, es una banda de inestabilidad ``pura''. 
Lo que implica este resultado es extremadamente 
importante: cualquier informaci\'on que pueda inferirse a trav\'es de an\'alisis 
astrosismol\'ogicos acerca de la estructura interna de las ZZ Ceti   
es aplicable tambi\'en a las enanas blancas DA no variables en general. 

\begin{figure}[!ht]
  \centering
  \includegraphics[width=0.45\textwidth]{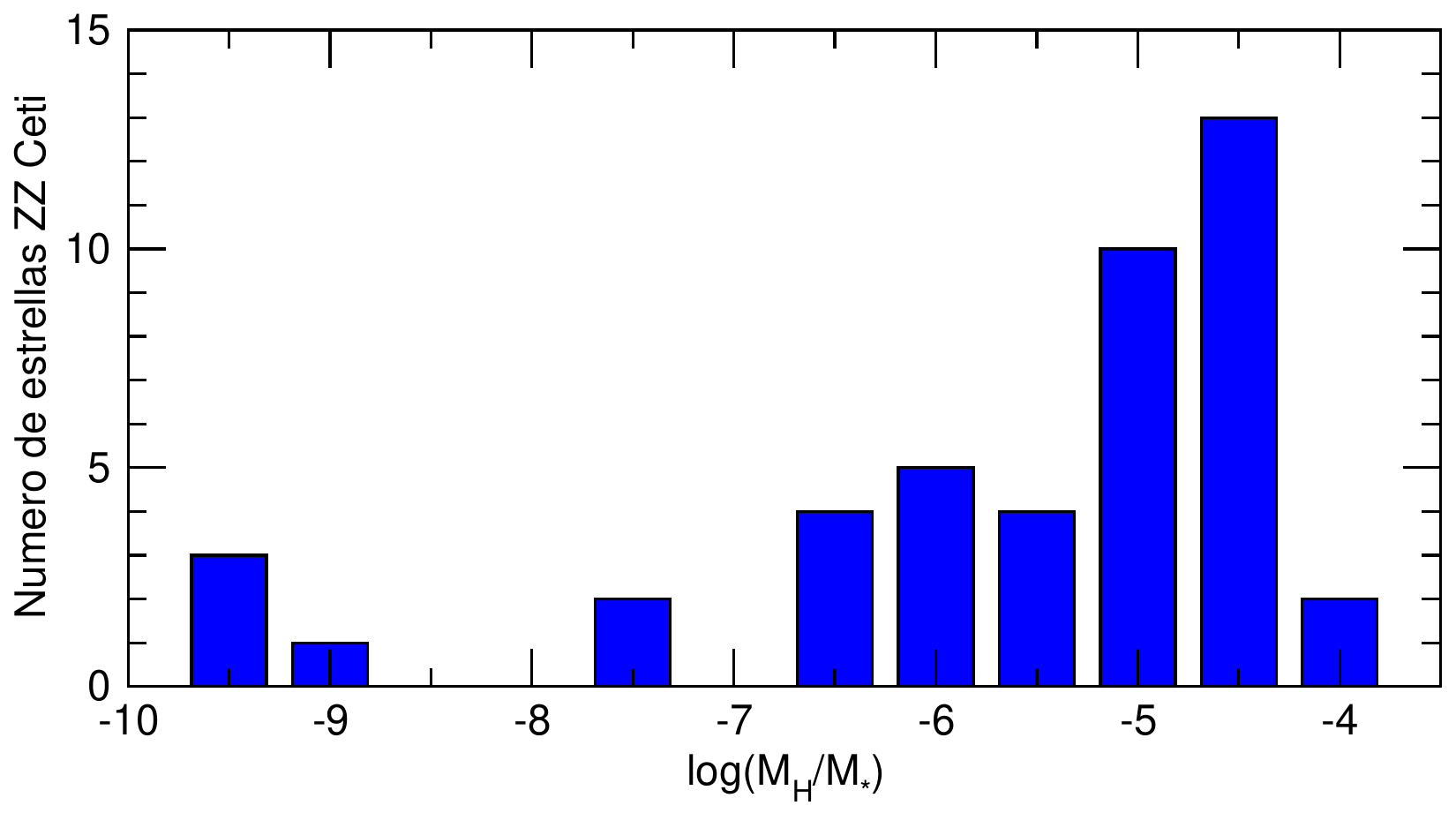}
  \caption{Histograma mostrando la distribuci\'on de espesores de la envoltura de H
para un ensamble de 44 estrellas ZZ Ceti brillantes \citep[ver][para mas detalles]
{2012MNRAS.420.1462R}. 
Este resultado podr\'ia tener validez para todas las estrellas enanas 
blancas DA en general.}
  \label{fig_04}
\end{figure}

Con respecto a los an\'alisis astrosismol\'ogicos de estrellas ZZ Ceti, hasta el momento
se han aplicado dos metodolog\'ias. Una de ellas  consiste en el empleo de 
modelos estelares simples y est\'aticos (es decir, que no provienen de c\'alculos 
evolutivos), caracterizados por perfiles qu\'imicos parametrizados {\em ad hoc}. El otro 
enfoque, el cual es el adoptado por el Grupo de Evoluci\'on Estelar y Pulsaciones 
de La Plata, emplea modelos resultantes de la evoluci\'on completa de las estrellas
progenitoras, desde la ZAMS ({\em Zero Age Main Sequence}) hasta la etapa de 
enana blanca. Estos modelos son obtenidos con el c\'odigo evolutivo {\tt LPCODE}.
Nuestro enfoque involucra el uso de ingredientes f\'isicos 
\citep[ecuaci\'on de estado, opacidades, eventos de mezcla, difusi\'on de elementos, etc; 
ver][]{2005A&A...435..631A} lo mas detallados y actualizados posibles. 
Esto es particularmente relevante con respecto a la estructura 
qu\'imica resultante, que constituye un aspecto crucial para interpretar 
correctamente los patrones de per\'iodos observados en las estrellas ZZ Ceti. 
Empleando esta metodolog\'ia, nuestro grupo ha llevado a cabo el primer estudio 
astrosismol\'ogico detallado sobre un ensamble de 44 estrellas ZZ Ceti mediante el uso de
modelos completamente evolutivos de enanas blancas DA \citep{2012MNRAS.420.1462R}. 
Los espectros de per\'iodos han sido calculados empleando la versi\'on 
adiab\'atica del 
c\'odigo de pulsaciones {\tt LP-PUL} \citep{2006A&A...454..863C}. 
Con el fin de ampliar el espacio de par\'ametros, en dicho estudio se 
consideraron adicionalmente envolturas de H ($M_{\rm H}$) mas delgadas (generadas 
artificialmente) que las que predice la 
teor\'ia est\'andar de evoluci\'on de enanas blancas (envolturas ``can\'onicas'').
Con el fin de encontrar un modelo o soluci\'on 
astrosismol\'ogica\footnote{Un modelo de enana blanca 
que reproduce con alto grado de precisi\'on los per\'iodos 
observados de la estrella en estudio.} para cada estrella de la muestra, 
hemos buscado los m\'inimos de una funci\'on de calidad $\chi^2(M_{\star},  
M_{\rm H}, T_{\rm eff})$ de la forma $\chi^2= \frac{1}{N} \sum_{i=1}^N 
[\Pi_k^{\rm th}-\Pi_i^{\rm obs}]^2$, donde $\Pi_k^{\rm th}$ y $\Pi_i^{\rm obs}$ 
son los per\'iodos te\'oricos y 
los per\'iodos observados, respectivamente, y $N$ es el n\'umero de per\'iodos observados.
Uno de nuestros objetos de estudio fue la estrella ZZ Ceti 
arquet\'ipica, G117$-$B15A, para la cual obtuvimos 
una \'unica soluci\'on astrosismol\'ogica, de esta 
forma eliminando la ambig\"uedad de larga data reportada en trabajos previos. 
Uno de los resultados mas importantes de este estudio sobre el ensamble de 44 estrellas 
es que existe un amplio rango de espesores de la envoltura de H (ver Figura \ref{fig_04})
para las estrellas ZZ Ceti analizadas.
La distribuci\'on de espesores exhibe una acumulaci\'on muy marcada para envolturas gruesas 
[$\log(M_{\rm H}/M_{\star}) \sim -4.5$]. Si bien estas envolturas son gruesas,  ellas son 
todav\'ia mas delgadas que las envolturas ``can\'onicas''. Tambi\'en es visible 
otro pico, mucho menos pronunciado,
correspondiente a envolturas muy delgadas [$\log(M_{\rm H}/M_{\star}) \sim -9.5$].  
Similares resultados 
han sido obtenidos posteriormente sobre una muestra de ZZ Cetis masivas 
\citep{2013ApJ...779...58R}. En resumen, estos resultados presentan un verdadero 
desaf\'io para los escenarios evolutivos est\'andar de 
formaci\'on de enanas blancas DA, los cuales predicen que estos objetos se forman 
\'unicamente con envolturas gruesas de H.

A pesar de los importantes resultados alcanzados usando esta metodolog\'ia, 
existen a\'un incertezas importantes conectadas con la evoluci\'on previa de 
los progenitores de las enanas blancas, tales como la cantidad precisa de mezcla extra 
({\em overshooting}) durante la etapa de combusti\'on central de He, el valor exacto de
la tasa de reacciones nucleares $^{12}$C($\alpha, \gamma)^{16}$O, la eficiencia de
la difusi\'on de elementos durante el enfriamiento de la enana blanca, etc. En este
sentido, en nuestros grupo estamos dando los primeros pasos para evaluar 
el impacto de dichas incertezas sobre las propiedades de los modelos astrosismol\'ogicos 
de las estrellas ZZ Ceti (De Ger\'onimo et al. 2015, en preparaci\'on).

\section{Pulsaciones en estrellas ELMV}
\label{ELMVS}

La distribuci\'on de masas de las enanas blancas exhibe una acumulaci\'on
para valores $M_{\star} \lesssim 0.45 M_{\sun}$. El origen de estas enanas blancas de baja
 masa, las cuales son del tipo espectral DA (es decir, poseen atm\'osferas ricas en H), 
estar\'ia ligado a la evoluci\'on binaria, en la cual una estrella de 
$\sim 1M_{\sun}$ con fuertes p\'erdidas de masa en la fase de RGB ({\em Red Giant Branch}) 
evitar\'ia el flash de He \citep{2013A&A...557A..19A}. Como resultado, 
estos objetos deber\'ian  tener n\'ucleos de He, a diferencia de las enanas blancas 
de masa promedio, las cuales poseen probablemente n\'ucleos hechos de C y O. 

En general, los progenitores de las enanas blancas de baja masa experimentan m\'ultiples 
flashes de H, lo cual conduce a una reducci\'on importante en el espesor de la envoltura 
de H. Sin embargo, en el caso de las llamadas enanas blancas ELM 
({\em Extremely Low Mass}), caracterizadas por masas $M_{\star}\lesssim 0.18-0.20 M_{\odot}$,
los progenitores no experimentan flashes, lo cual resulta en envolturas muy gruesas.
La presencia de envolturas muy gruesas en las ELMs es la raz\'on por la cual, 
a diferencia de lo que sucede en las enanas blancas en general, la combusti\'on
nuclear estable del H a trav\'es de la cadena $pp$ es muy importante 
y controla su evoluci\'on\footnote{Un fen\'omeno 
similar se verifica en enanas blancas de masa promedio provenientes de progenitores 
de muy baja metalicidad, $Z \lesssim 0.001$; ver \cite{2013ApJ...775L..22M} y 
Camisassa et al. (2015) en este Bolet\'in.}.
Esto conduce a  edades de  enfriamiento enormes en este tipo de estrellas 
($\tau \sim 10^9$ a\~nos) en comparaci\'on con aquellas con masas 
$M_{\star} \gtrsim 0.18-0.20 M_{\star}$ ($\tau \sim 10^7$ a\~nos). Este resultado se 
conoce como ``dicotom\'ia de edades'' \citep{2001MNRAS.323..471A,2013A&A...557A..19A}.

\cite{2010ApJ...718..441S} fueron los primeros en sugerir que las enanas blancas ELM
deber\'ian pulsar en modos $g$ de la misma naturaleza que las ZZ Ceti. Dicha 
predicci\'on fue confirmada  por \cite{2012ApJ...750L..28H}, quienes descubrieron la primer
enana blanca ELM variable (ELMV). El origen de las pulsaciones fue explicado en detalle 
por \cite{2012A&A...547A..96C}, quienes llevaron a cabo detallados c\'alculos de estabilidad
pulsacional. Estos autores encontraron que una gran cantidad de modos $g$, con 
per\'iodos similares a los 
observados, son excitados a trav\'es del mecanismo $\kappa-\gamma$ actuando en la 
zona de ionizaci\'on parcial del H  en modelos de enanas 
blancas ELM. Estos resultados fueron
mas tarde confirmados por los c\'alculos independientes de \cite{2013ApJ...762...57V}. 

\begin{figure}[!ht]
  \centering
  \includegraphics[width=0.45\textwidth]{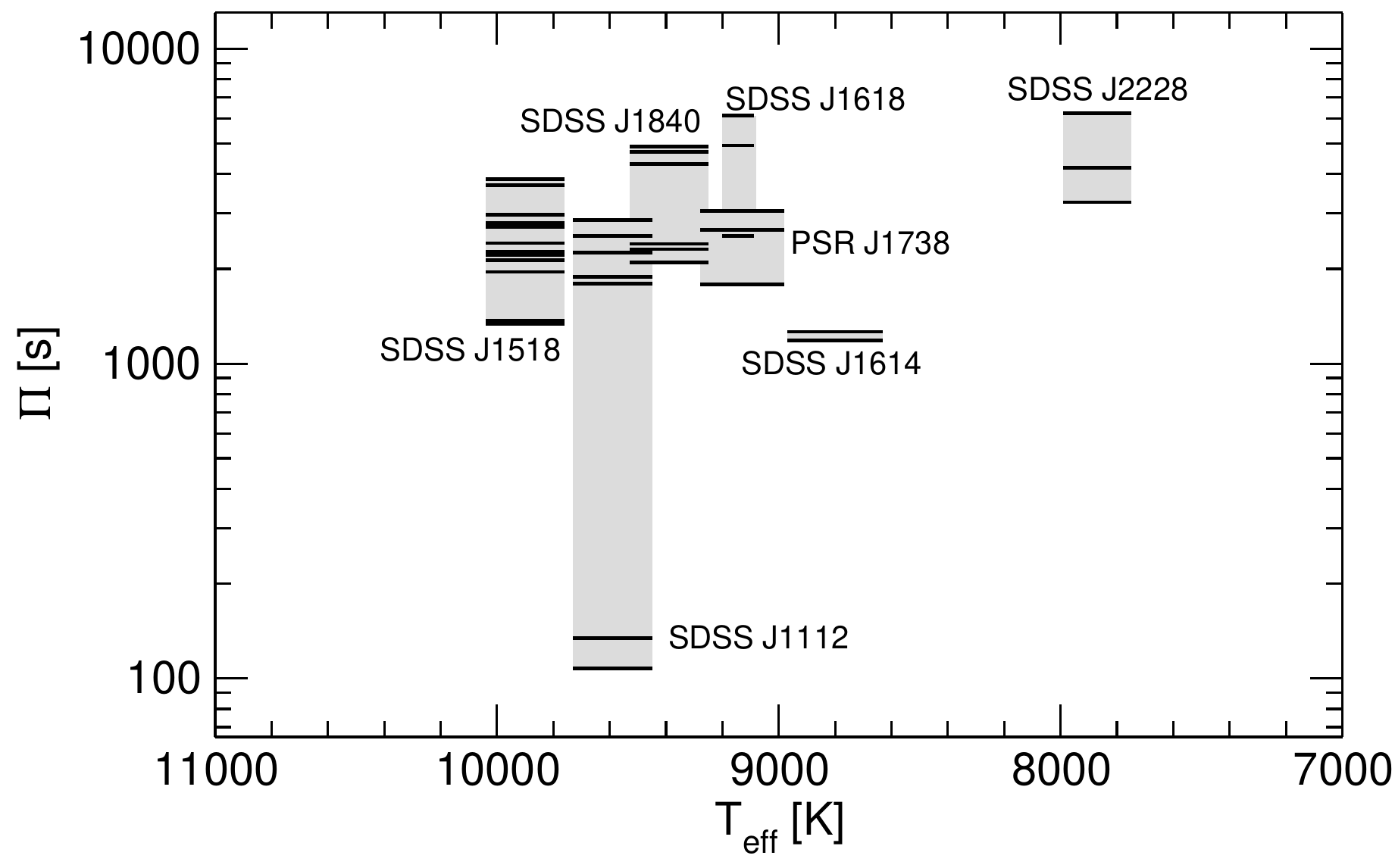}
  \caption{Per\'iodos de pulsaci\'on de las 7 estrellas ELMVs conocidas en t\'erminos
de la temperatura efectiva.}
  \label{fig_05}
\end{figure}

Nuevas observaciones fotom\'etricas han llevado a la detecci\'on de otras estrellas ELMVs
\citep[][Bell et al., 2014\footnote{19th European Workshop on White Dwarfs, Montréal, Canada.}, 
2015, en preparaci\'on]{2013ApJ...765..102H,2013MNRAS.436.3573H,2015MNRAS.446L..26K}, con lo cual 
actualmente se conocen 7 objetos de este peque\~no grupo de enanas blancas pulsantes.
Las estrellas ELMVs exhiben en general per\'iodos muy largos. En la Figura \ref{fig_05} 
se muestran los per\'iodos detectados en las 7 estrellas conocidas. Los per\'iodos 
largos pueden explicarse con comodidad como debidos a modos $g$ con grado 
arm\'onico $\ell= 1$ o $\ell= 2$.

Es interesante notar que la estrella SDSS J1112 exhibe dos per\'iodos muy cortos, 
de $\sim 108$ s y 
$\sim 134$ s, adem\'as de per\'iodos largos. Estos per\'iodos an\'omalos podr\'ian 
estar asociados a modos $p$ de 
bajo orden radial $k$, o inclusive a modos radiales. Sin embargo,
si la temperatura efectiva y la gravedad de SDSS J1112 son correctas, los modelos
de enanas blancas ELM no son capaces de reproducir esos per\'iodos 
con modos $p$ y/o modos radiales \citep{2014A&A...569A.106C}. Otra posibilidad 
es que se trate de modos $g$ de bajo orden radial. Esta \'ultima opci\'on, sin 
embargo, es dif\'icil de concebir por el hecho de que los resultados te\'oricos 
predicen inestabilidad para per\'iodos mas largos, $\Pi \gtrsim 1100$ s
\citep{2012A&A...547A..96C,2013ApJ...762...57V}.

\begin{figure}[!ht]
  \centering
  \includegraphics[width=0.45\textwidth]{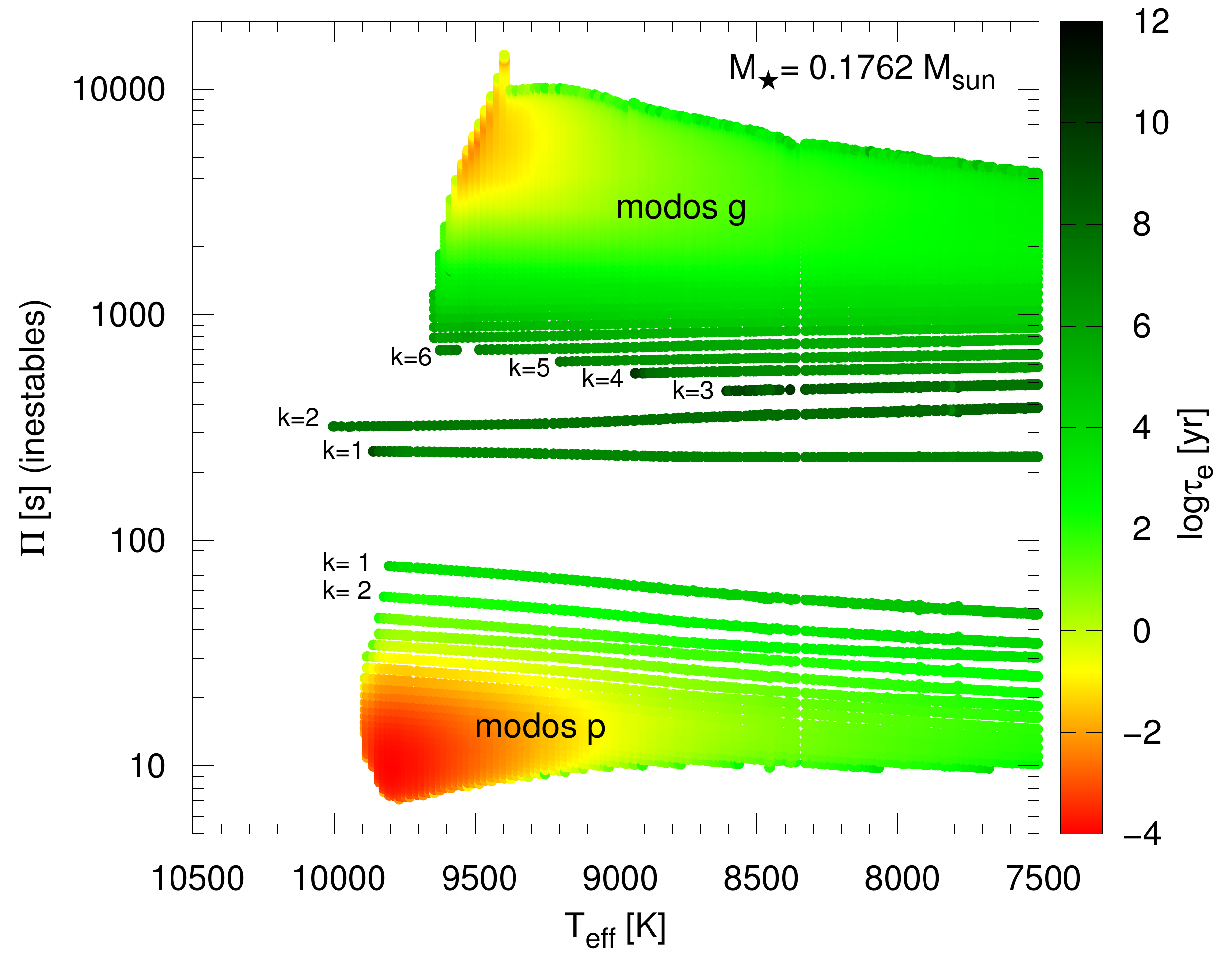}
  \caption{Per\'iodos de modos dipolares ($\ell= 1$) 
inestables  en funci\'on de la temperatura
efectiva, correspondientes a un modelo de enana
blanca ELM con masa $M_{\star} = 0.1762 M_{\sun}$. La escala de la 
izquierda con el c\'odigo de colores corresponde al {\em e-folding time}
(la escala de tiempo de 
crecimiento de las amplitudes).}
  \label{fig_06}
\end{figure}

\begin{figure}[!ht]
  \centering
  \includegraphics[width=0.45\textwidth]{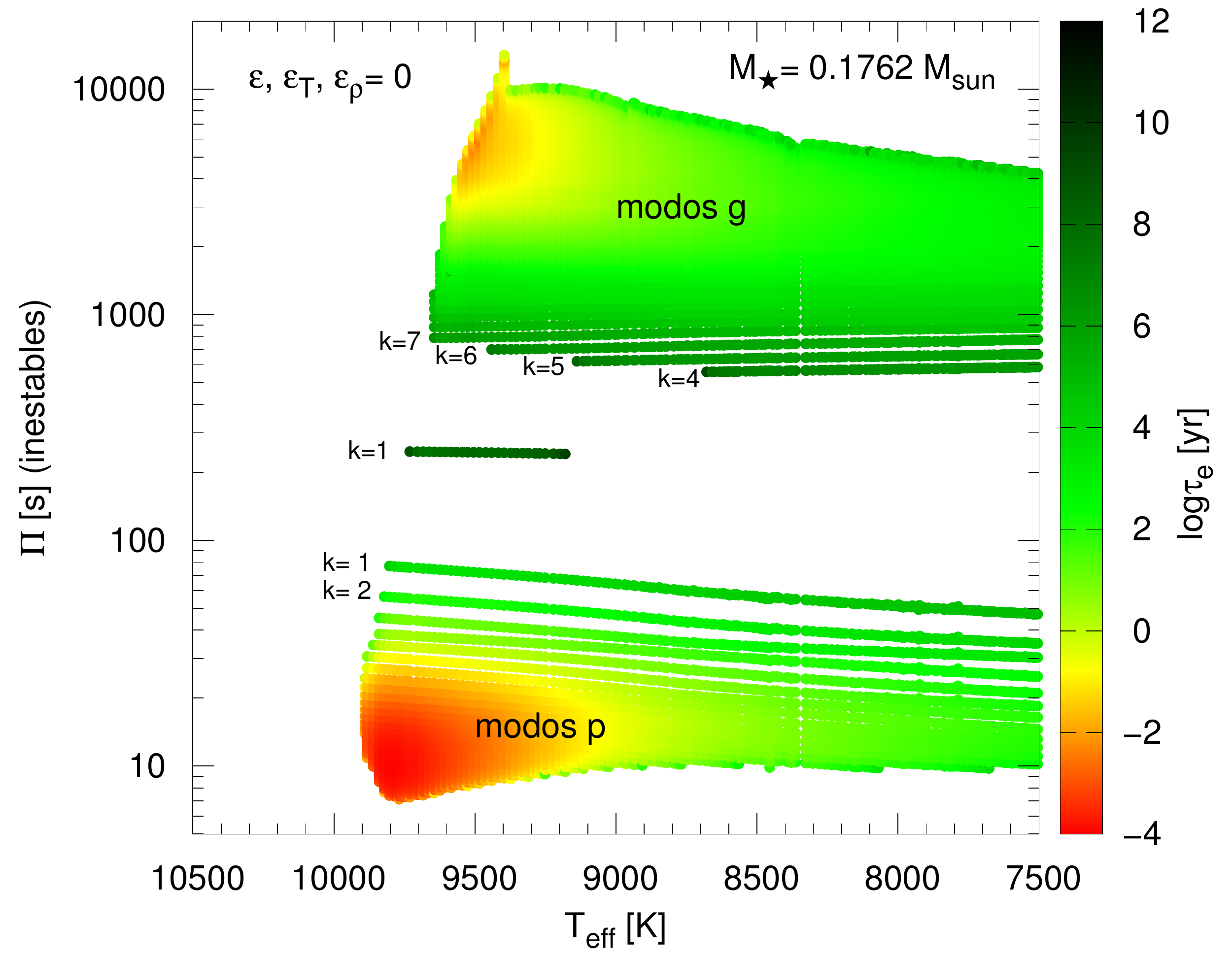}
  \caption{Igual que  la Figura \ref{fig_06}, pero para el caso en que el 
mecanismo $\varepsilon$ es suprimido 
($\varepsilon= \varepsilon_{\rm T}= \varepsilon_{\rho}= 0$).}
  \label{fig_07}
\end{figure}

Dado que, como mencionamos, estas estrellas poseen envolturas muy gruesas de H,
ellas son capaces de sostener combusti\'on de H en forma estable, lo que provoca un 
retardo importante en su enfriamiento. En el Grupo de Evoluci\'on Estelar y 
Pulsaciones de La Plata decidimos considerar la posibilidad de que esos modos de corto 
per\'iodo sean modos $g$ de bajo orden $k$ excitados por el mecanismo $\varepsilon$ 
actuando en la capa de H en combusti\'on. Para tal fin, hemos llevado a cabo c\'alculos 
de estabilidad pulsacional sobre una grilla de secuencias evolutivas de enanas blancas 
ELM provenientes de la evoluci\'on binaria y publicadas en \cite{2013A&A...557A..19A}.
Los c\'alculos de estabilidad han sido realizados empleando la 
versi\'on no adiab\'atica del c\'odigo de pulsaciones {\tt LP-PUL} 
\citep{2006A&A...458..259C}.
En la Figura \ref{fig_06} graficamos 
los per\'iodos de  modos inestables con grado arm\'onico $\ell= 1$ en t\'erminos de 
la temperatura efectiva, correspondientes a un modelo
de enana blanca ELM con masa $M_{\star} = 0.1762 M_{\sun}$. El c\'odigo de colores 
(escala de la izquierda) indica el 
valor del logaritmo de la escala de tiempo de crecimiento de las amplitudes 
({\em e-folding times}, $\tau_e$) de cada modo inestable. La figura 
muestra un espectro de modos $g$ y $p$ excitados por 
el mecanismo $\kappa-\gamma$ 
actuando en la regi\'on de ionizaci\'on parcial del H. En todos los casos,
la escala de tiempo de crecimiento de la amplitud de los modos  
es entre 10 y 100 veces mas corta que la evolutiva, con lo cual los modos excitados
tienen tiempo suficiente para alcanzar amplitudes observables. Si se suprime 
el mecanismo $\varepsilon$ en los c\'alculos de estabilidad pulsacional, 
los modos $g$ con $k= 1, 2, 3$ y $4$ ya no son inestables (en ciertos rangos de 
$T_{\rm eff}$), como queda claramente evidenciado en la Figura \ref{fig_07}. 
Esto implica que, en efecto, el mecanismo $\varepsilon$
es capaz de desestabilizar modos $g$ de bajo orden radial con per\'iodos cortos 
similares a los exhibidos por SDSS J1112. Un acuerdo mas cercano con los per\'iodos 
observados puede lograrse considerando modos cuadrupolares ($\ell= 2$), como
se detalla en el art\'iculo de \cite{2014ApJ...793L..17C}. Si la realidad 
de los modos de corto per\'iodo observados en esta estrella es confirmada, 
esta podr\'ia ser la primera demostraci\'on de la existencia de 
combusti\'on nuclear en enanas blancas fr\'ias.

\begin{acknowledgement}
Quisiera agradecer a los miembros del Comit\'e Cient\'ifico de la $57^{\circ}$ 
Reuni\'on de la Asociaci\'on Argentina de Astronom\'ia por la amable invitaci\'on 
para presentar este informe.
\end{acknowledgement}


\bibliographystyle{baaa}
\small
\bibliography{corsico}
 
\end{document}